\documentclass[twocolumn,showpacs,amsmath,prl]{revtex4}

\usepackage{graphicx}

\newcommand{\ve}[1]{\mathbf{#1}}

\begin{document}


\title{Excited State Properties of Organic Semiconductors: Breakdown of the Tamm-Dancoff Approximation}

\author{Peter Puschnig}
 \email{peter.puschnig@uni-graz.at}
 \altaffiliation{Present address: Institut f\"ur Physik, Karl-Franzens-Universit\"at Graz, Universit\"atsplatz 5, 8010 Graz, Austria}
\author{Christian Meisenbichler}
\author{Claudia Ambrosch-Draxl}
\altaffiliation{Present address: Institut f\"ur  Physik, Humboldt-Universit\"at zu Berlin, Newtonstra{\ss}e 15, 12489 Berlin, Germany}
\affiliation{%
Chair for Atomistic Modelling and Design of Materials, Montanuniversit\"at Leoben,
Franz-Josef-Stra\ss e 18, A--8700 Leoben, Austria.
}%

\date{\today}
             
\begin{abstract}
The solution of the Bethe-Salpeter equation within the framework of many-body
perturbation theory has turned out to be a benchmark for ab-initio calculations of
optical properties of semiconductors and insulators. Commonly, however, the
coupling between the resonant and anti-resonant excitations is neglected which is
referred to as the Tamm-Dancoff approximation (TDA). This is well justified in
cases where the exciton binding energy is much smaller than the band gap. Here, we
report on the optical properties of a representative series of organic
semiconductors where we find the TDA to no longer hold. We observe an increase of
the exciton binding energy of up to 44\% thereby improving the agreement with experiment.
\end{abstract}

\pacs{71.10.-w, 71.15.Qe, 71.35.-y, 78.40.Me}
\maketitle



\paragraph{Preamble.} This work has been conducted already in the years 2006--2007 but was not published at that time \cite{Denver2007}. In the meantime several papers have come out which report similar observations regarding the breakdown of the Tamm-Danncoff approximation (TDA), most notably  a paper by Gr\"uning and co-workers on the exciton binding energies in organic molecules and carbon nanotubes \cite{Gruning2009}. The role of the TDA for excited states of biological chromophores has also been assessed in the paper by Ma and coworkers \cite{Ma2009}. Nevertheless, the results reported in this manuscript may prove useful for the experts in the field, which motivated us to make them available now. Note that the remaining text of the paper, all figures and tables are kept unaltered from  their 2007 version.

Within the framework of many-body perturbation theory the linear response to an optical perturbation is
expressed in terms of the equation of motion for the electron-hole (e-h) two-particle
Green's function, the so-called Bethe-Salpeter equation (BSE).  The solution of the BSE 
for the e-h correlation function represents a systematic first-principles approach for the calculation of 
optical absorption spectra including excitonic effects \cite{Onida2002}. Solutions of the
BSE in an \emph{ab-initio} framework 
have shown that e-h interactions are indeed important in order to correctly account for
quantitative (oscillator strengths) as well as qualitative (bound excitons) features of
optical spectra of semiconductors and insulators. This is true for inorganic
\cite{Albrecht1997,Benedict98,Rohlfing98,Arnaud01,Puschnig2002a}
as well as for organic semiconductors 
\cite{Rohlfing99,vanderHorst99,Ruini02,Bussi2002,Puschnig2002b,Tiago03,Hummer2004,Hummer2005a}.
The exciton binding energy (BE) is a central quantity in the photophysics of
these materials since it is intimately related to the probability of radiative
emission/absorption and electric-field induced generation of free charge carriers. 
Indeed, ab-initio BSE results have considerably contributed to the longstanding debate
about the nature of the energetically lowest optical excitations  in  organic
semiconductors \cite{Rohlfing99,vanderHorst99,Ruini02,Bussi2002,Puschnig2002b,Hummer2004,Tiago03}. 
However, approximations in the state-of-the-art
BSE approach, which have been proven to be valid for bulk inorganic semiconductors such
as Si or GaAs \cite{Onida2002}, do not \emph{a priori} hold for the highly anisotropic organic semiconductors. 
Transitions at positive (resonant part) and negative (anti-resonant part) frequencies
are assumed to be decoupled leading to the Tamm-Dancoff approximation (TDA) \cite{Benedict98}. 
This approximation is well justified, whenever exciton binding
energies are much smaller than the band gap and it has been shown to have no influence on the
optical absorption spectrum of bulk Si \cite{Olevano01,Onida2002} while 
the coupling between the resonant and anti-resonant terms becomes relevant for the
electron energy loss spectrum even in this case \cite{Olevano01}.

In this Letter, we demonstrate that going beyond the TDA, i.~e., taking into account the
full matrix structure of the BSE, becomes indeed necessary also for the optical absorption
when treating organic semiconductors where exciton binding energies amount up to 20\% of the band gap.
In order to solve the full BSE matrix problem numerically we adapt the time-evolution scheme
proposed for the TDA-BSE \cite{Schmidt03}. This has the advantage that it (i) allows for well-converged
spectra with respect to $\ve{k}$ grid and number of included bands, (ii) poses no technical problems
due to the non-Hermitian character of the full BSE matrix, and (iii) finally enables an 
efficient parallelization of the numerical computations.  

Starting from the BSE in its integral form it can be transformed into
a matrix eigenvalue equation by expanding all quantities in terms of
single-particle electron and hole states $\psi_{c \ve{k}}$ and $\psi_{v \ve{k}}$, respectively \cite{Rohlfing00}.
From the resulting matrix eigenvalue problem the excitations energies (eigenvalues)
as well as the electron-hole coupling coefficients (eigenvectors) can be obtained.
The matrix structure of this effective electron-hole Hamiltonian $H$ is given by the 
following expression \cite{Olevano01}
\begin{equation}
\label{eq:matrixstructure}
H = \left( 
\begin{array}{cc}
  R   & C      \\
 -C^* & -R^*
\end{array}
     \right)			     	
\end{equation}
where the diagonal blocks $R$ and the coupling blocks $C$, respectively, have been defined 
according to 
\begin{eqnarray}
\label{eq:defineexchamilton}
  R & = & 
 (E_{c_1 \ve{k}_1} - E_{v_1 \ve{k}_1})
 \delta_{v_1 v_2} \delta_{c_1 c_2} \delta_{\ve{k}_1 \ve{k}_2}
 + i \; \Xi_{v_1 c_1 \ve{k}_1, v_2 c_2 \ve{k}_2}  \nonumber \\
  C & = &
 + i \; \Xi_{v_1 c_1 \ve{k}_1, c_2 v_2 \ve{k}_2}  .
\end{eqnarray}
Here, $E_{c_1 \ve{k}_1}$ and $E_{v_1 \ve{k}_1}$, respectively, denote the
electron and hole quasi-particle energies, while $\Xi_{v_1 c_1 \ve{k}_1, v_2 c_2 \ve{k}_2}$
represents the electron-hole interaction kernel given by the sum of the bare
exchange and the screened direct interaction \cite{Onida2002}. Note that
the diagonal blocks contain the difference in the quasi-particle energy
and are therefore greater than the single-particle band gap $E_g$. Hence, the coupling
matrices $C$ can be neglected when electron-hole interactions are small compared to $E_g$, 
which results in the TDA.
Employing the TDA  reduces the size of the eigenvalue problem of the full
BSE matrix $H$ by a factor of 2 and leaves an eigenvalue problem for the Hermitian matrix $R$.

In order to go beyond the TDA we start with a general expression for the frequency dependent  
macroscopic polarizability tensor $\alpha^{M}_{ij}$  \cite{Onida2002,Schmidt03}
where $i$ and $j$ stand for the Cartesian coordinates $x$, $y$, $z$
\begin{eqnarray}
\label{eq:polarizability}
\alpha^{M}_{ij}  \left( \omega \right) & = &
  \frac{8 \pi}{\Omega} \sum_{\lambda \lambda'}
  \langle \mu^i | A^{\lambda} \rangle 
   S^{-1}_{\lambda \lambda'} 
  \langle A^{\lambda'} | \mu^j \rangle \nonumber \\
  & \times &
  \left[ 
           \frac{1}{E^{\lambda} - \omega - i \delta}
         + \frac{1}{E^{\lambda} + \omega + i \delta}
  \right].
\end{eqnarray}
$E^{\lambda}$ and $| A^{\lambda} \rangle$ denote the eigenvalues
and eigenvectors of the full Hamiltonian, $S^{-1}_{\lambda \lambda'}$ is the inverse
of the overlap matrix $S_{\lambda \lambda' } = \langle  A^{\lambda'} | A^{\lambda} \rangle$,
$\delta$ is a broadening parameter,
and $| \mu^j \rangle$ is given by the optical matrix elements normalized by the
quasi-particle energy differences 
\begin{equation}
\mu^{j}_{vc\ve{k}} = 
\frac{\langle v \ve{k} | \nabla_j | c \ve{k} \rangle}{E_{c \ve{k}} - E_{v \ve{k}}}.
\end{equation}
The scalar products $\langle . | . \rangle$ in Eq.~(\ref{eq:polarizability}) 
involve summations over all valence, conduction bands, and $\ve{k}$ points $(vc\ve{k})$
of the full matrix structure. A direct evaluation of Eq.~(\ref{eq:polarizability}) by diagonalizing
$H$ is complicated by the fact that $H$ is non-Hermitian which makes the numerical
solution of the eigenvalue problem non-trivial. Moreover, Eq.~(\ref{eq:polarizability}) also involves the inverse
of the overlap matrix, hence another computationally demanding task would be required. Instead we follow a 
route proposed by Schmidt et al.~for the TDA-BSE \cite{Schmidt03} in which 
the macroscopic polarizability is obtained without requiring the eigenvalues of $H$ explicitly.
It can be shown that the polarizability can be equivalently expressed as
\begin{equation}
\label{eq:polarizability2}
\alpha^{M}_{ij} \left( \omega \right) =
  \frac{8 \pi}{\Omega} i \int_0^\infty dt
  e^{i \left( \omega + i \delta \right) t}
  \left[
     \langle \mu^i | \xi^j \left( t \right) \rangle
    -\langle \mu^i | \xi^j \left( t \right) \rangle^*
  \right].
\end{equation}
Here, we have introduced the time dependent vectors $| \xi^j \left( t \right) \rangle$ whose
time evolution is governed by the unitary transformation
\begin{equation}
\label{eq:xidefinition}
| \xi^j \left( t \right) \rangle = e^{- i \hat{H} t} | \mu^j \rangle .
\end{equation}
The exponential of the Hamiltonian is defined via the spectral theorem as
\begin{equation}
\label{eq:spectral}
 e^{- i \hat{H}   t} = \sum_{\lambda \lambda'} 
 e^{- i E^\lambda t} | A^{\lambda} \rangle 
                       S^{-1}_{\lambda \lambda'}
                      \langle A^{\lambda'} |. 
\end{equation}
Inserting Eqs.~(\ref{eq:xidefinition}) and (\ref{eq:spectral}) into (\ref{eq:polarizability2})
it is straight forward to demonstrate that both equations for the polarizability,  
i.~e.~Eqs.~(\ref{eq:polarizability}) and (\ref{eq:polarizability2}), are mathematically equivalent.
Inspection of Eq.~(\ref{eq:polarizability2}) reveals that $\alpha^{M}_{ij} \left( \omega \right)$
is obtained from the Fourier transform of the time-dependent quantity in square brackets 
evolving in time according to the Schr\"odinger equation
\begin{equation}
\label{eq:timeevolution}
i \frac{d}{dt}| \xi^j \left( t \right) \rangle = \hat{H} | \xi^j \left( t \right) \rangle .
\end{equation}
Setting the initial value  $\xi^j \left( 0 \right) \rangle = \mu^j$ this 
is equivalent to the definition (\ref{eq:xidefinition}). Eq.~(\ref{eq:timeevolution})
can be integrated numerically, e.~g.~using an explicit scheme where 
the time step $\delta t$ for the numerical integration is given by the usual
stability criterion \cite{Schmidt03}.


\begin{figure}[tb]
\begin{center}
\includegraphics[width=\columnwidth]{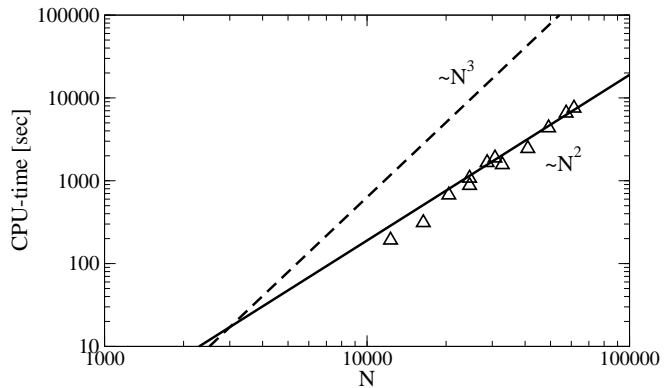}
\end{center}
\vspace{-5mm}
\caption[CPU time]
{CPU time for the time evolution algorithm as a function of matrix size $N$ scaling
as $N^2$. For comparison the $N^3$ behavior of a matrix diagonalization scheme is indicated.}
\label{fig:cpu}
\end{figure}

All calculations presented in this Letter have been obtained by 
utilizing the full-potential linearized augmented plane wave plus local orbitals method (FP-LAPW) 
as realized in the WIEN2K code \cite{wien2k}. The implementation of the TDA-BSE scheme within the
FP-LAPW method has been described elsewhere \cite{Puschnig2002a,exciting}. Convergence with respect to
the $\ve{k}$ grid and number of valence and conduction bands has been checked. For all spectra we have chosen
a broadening of $0.2$ eV which typically requires time steps of $\Delta  t \approx 0.01$ fs and 
around 3000 time steps within the time evolution scheme. Fig.~\ref{fig:cpu} compares the
CPU time of the present time evolution method with a traditional matrix diagonalization scheme as a function
of the matrix size $N=2 N_\ve{k} N_v  N_c$, where $N_\ve{k}$ denotes the number of $\ve{k}$ points,
and $N_v$ and $N_c$ are the number of valence and conduction states, respectively. The factor of 
2 arises from the doubling of the matrix size due to the inclusion of the coupling blocks $C$.
Clearly, one observes an $N^2$ scaling for the time evolution scheme where the cross-over with
the $N^3$ behavior of the diagonalization scheme already takes place at a moderate matrix size of around 2000.
We note that the time evolution scheme allows for well-converged spectra at reasonable computational
effort and the matrix multiplications necessary for the integration of Eq.~(\ref{eq:timeevolution})
makes efficient parallelization of the code possible.


We have benchmarked our approach with bulk Si using a $8 \times 8 \times 8$ $\ve{k}$ mesh,
4 valence and 15 conduction states. Thus a matrix size of $\approx 60000$ was sufficient 
in order to obtain converged spectra for the loss function. As has already been noted earlier
\cite{Olevano01} the inclusion of the resonant-antiresonant coupling terms $C$ has
negligible effects on the optical spectra. We note that going beyond the TDA
reduces the static dielectric constant of Si by only 3.5\%. On the other hand,
taking into account the full matrix structure of the BSE has sizable effects on
the  electron loss function thereby improving the experimental agreement
\cite{Olevano01}.

\begin{figure}[ptb]
\begin{center}
\includegraphics[width=\columnwidth]{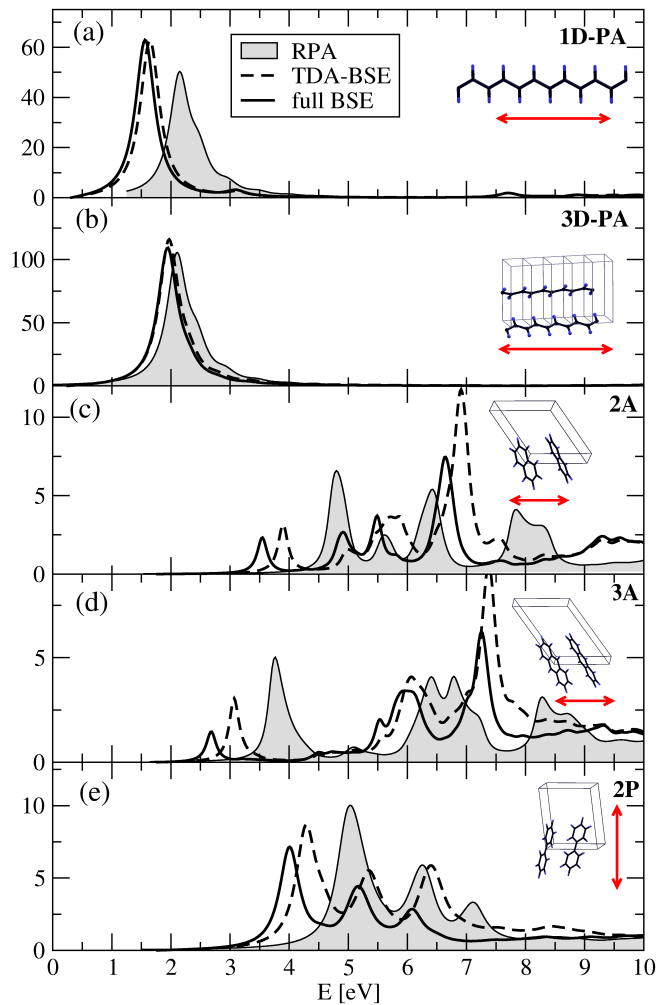}
\end{center}
\vspace{-5mm}
\caption[Main result]
{Imaginary part of the dielectric function calculated within the RPA (shaded area), the
TDA-BSE (dashed line) and the full BSE (continuous line) for (a) isolated polyacetylene chain,
(b) three-dimensional arrangement of polyacetylene chains, molecular crystals of (c) naphthalene, 
(d) anthracene, and (e) biphenyl. The arrows indicate the orientation of the polarization vector
of the exciting light wave.}
\label{fig:main}
\end{figure}

\begin{table}[ptb]
\caption{Exciton binding energies obtained with the TDA, $E_b^\textrm{TDA}$, and by 
including the coupling terms in the BSE, $E_b^\textrm{fBSE}$. For comparison the static dielectric constants
$\epsilon$ in $x$, $y$ and $z$ directions are also given.}
\label{tab:results}
{\begin{tabular}{ccccccc} \hline \hline
material    & $\epsilon_x$ &  $\epsilon_y$ & $\epsilon_z$ & $E_b^\textrm{TDA}$ [eV] & $E_b^\textrm{fBSE}$ [eV] & $\Delta E_b$ [eV] \\ \hline
1D-PA       & 1.9        & 1.5    & 33.0              & 0.50               & 0.60         &   0.10      \\ 
3D-PA       & 3.0        & 2.6    & 33.0              & 0.13               & 0.15         &   0.02           \\   
2A          & 2.8        & 3.8    &  \hphantom{0}4.9  & 0.90               & 1.30         &   0.40           \\ 
3A          & 3.1        & 4.1    &  \hphantom{0}6.3  & 0.70               & 1.00         &   0.30           \\     
2P          & 3.1        & 3.5    &  \hphantom{0}4.6  & 0.74               & 1.02         &   0.28           \\  
\hline \hline        
\end{tabular}}
\end{table}

\begin{figure}[ptb]
\begin{center}
\includegraphics[width=\columnwidth]{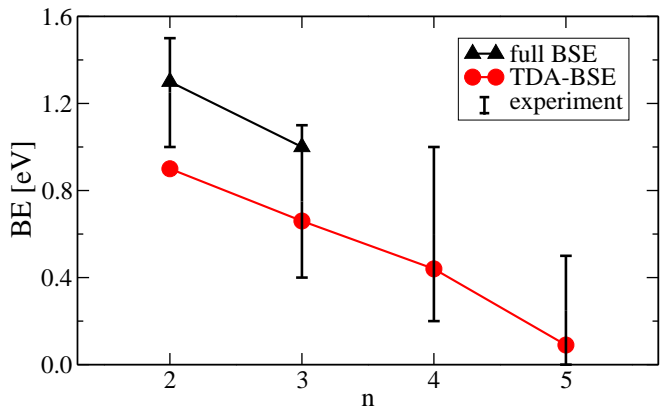}
\end{center}
\vspace{-5mm}
\caption[Comparison]
{Exciton binding energies for the oligoacene series from naphthalene ($n=2$) to
pentacene ($n=5$). Calculated values within the TDA and using the full BSE matrix
structure are compared with a range of experimental data as compiled in Ref.~\cite{Hummer2005a}.}
\label{fig:comparison}
\end{figure}

Bulk Si exhibits a high dielectric constant and therefore an
efficient screening of electron-hole interactions. Hence, the coupling matrices $C$
are small compared to the diagonal terms, which are in the order of $E_g$, and the 
effect on the optical absorption spectra remains marginal.
On the contrary, one can expect the resonant-antiresonant coupling to play a significant role in the 
excited state properties of organic semiconductors since exciton binding energies 
are in the range of 10--20\% of the band gap \cite{Hummer2006a}. 
We have chosen five prototypical organic semiconductors that cover the range from 
a three-dimensional arrangement of polymer chains with efficient electron-hole
screening to molecular crystals consisting of short oligomers where the electron-hole
wave function is spatially well-confined.
While for the former the exciton binding energies are comparably small and hence
effects of resonant-antiresonant coupling, the latter exhibits large excitonic effects
and enhancements of excitonic effects by going beyond the TDA are expected to
be strongest.   
Fig.~\ref{fig:main} displays the imaginary part of the dielectric
function for these five prototypical organic semiconductors using (i) the RPA, (ii) the
TDA-BSE and (iii) the full BSE. Panel (a) and (b), respectively, display
spectra for an isolated chain of trans-polyacetylene (1D-PA) and a
three-dimensional arrangement of PA chains (3D-PA) while panels (c)--(e) show
the results for molecular crystals consisting of small organic molecules,
namely naphthalene (2A), anthracene (3A) as well as biphenyl (2P). The exciton
BEs calculated within the TDA of the above mentioned organic semiconductors
are in the range between 0.13 and 0.9 eV \cite{Hummer2006a} as summarized in
Table~\ref{tab:results}.

Due to the structure of the full BSE matrix given in
Eq.~(\ref{eq:matrixstructure}) we can expect the coupling matrices to enhance
excitonic effects, i.~e.~shift oscillator strengths towards lower energy or
increase the exciton binding energies. This trend is indeed confirmed by our
results. For the polymer, both the 1D as well as the 3D case, the effect of
the TDA is moderate. Inclusion of the coupling matrices increases the BE from
0.5 to 0.6 eV for the 1D case while  it is only 0.02 eV for the 3D
arrangement of polymer chains.  For the molecular crystals, on the other hand,
we find a substantial enhancement of the exciton BE when going beyond the TDA.
In naphthalene (2A), the shortest oligomer under study, the BE increases by 0.4
eV. We emphasize that the resulting BE of 1.3 eV is in much better agreement
with experimental values than the previous TDA result \cite{Hummer2005a}. When
increasing the  length of the molecule, that is going from 2A to 3A, the
exciton BE decreases due to an enhanced screening of the electron-hole
interaction. This results in a TDA value of 0.7 eV while the full BSE
calculation gives 1.0 eV. Note that in the case of 3A both theoretical results
are within the large experimental error bars. A compilation of the present results for
the oligoacene series together with previous TDA results \cite{Hummer2005a} as
well as corresponding experimental data is given in in Fig.~\ref{fig:comparison}.
The decrease of the exciton BE with increasing chain length
reduces the resonant-antiresonant coupling effect. Hence,
we expect an improvement of the experimental agreement also for tetracene (4A)
and pentacene (5A), molecular crystals which are intermediate cases between 
the strongly localized situation of short molecules and the polymers.


In summary, we have found that the TDA does no longer hold in cases where the
exciton BE is large compared to the band gap and screening of the electron-hole
interaction is inefficient. Hence, the impact of resonant-antiresonant coupling on the
exciton binding energy is most pronounced for molecular crystals consisting of short
molecules for which we find an increase of up to 0.4 eV (44 \%) in the exciton binding energy.
We have utilized the numerically efficient time-evolution scheme for solving the
full BSE matrix including the coupling between positive and negative frequencies.
This approach avoids problems due to the non-Hermiticity of the full BSE matrix
and at the same time allows for well converged spectra and efficient parallelization of the code,
where we have found a $N^2$ scaling of the CPU time with matrix size $N$.
The present results should also prove valuable in the search for kernels to be used within
time-dependent density functional theory (TDDFT) that are capable of accounting for 
excitonic effects. In the past, only TDA-BSE results have been used to derive
\emph{ab-initio} linear response exchange-correlation kernels for TDDFT \cite{Sottile03,Marini2003b,Adragna2003}.
By demonstrating that the TDA substantially underestimates excitonic effects in 
organic semiconductors we also expect implications of our findings on the novel 
developments in the field of TDDFT.

\begin{acknowledgments}
We acknowledge the financial support from the Austrian Science Fund (project 16227-PHY 
and NFN research network S9714) and the funding from the EU RT network \emph{EXCITING},
contract number HPRN-CT-2002-00317.
\end{acknowledgments}


\end{document}